\documentclass[letterpaper]{article} 
\usepackage{aaai24}  
\usepackage{times}  
\usepackage{helvet}  
\usepackage{courier}  
\usepackage[hyphens]{url}  
\usepackage[cmyk]{xcolor}
\usepackage[skins, breakable]{tcolorbox}
\usepackage{graphicx} 
\usepackage{natbib}  
\usepackage{caption} 
\definecolor{lightgreen}{cmyk}{0.4,0,0.6,0}
\definecolor{darkgreen}{cmyk}{0.8,0,1,0.2}
\definecolor{lightorange}{cmyk}{0,0.2,0.7,0}
\definecolor{lightgrey}{cmyk}{0,0,0,0.1} 
\definecolor{darkgrey}{cmyk}{0,0,0,0.4}  
\newtcolorbox{greybox}{colback=lightgrey!5!white, colframe=lightgrey, breakable}
\newtcolorbox{darkgreybox}{colback=darkgrey!20!white, colframe=darkgrey, breakable}
\newtcolorbox{greenbox}{colback=lightgreen!5!white, colframe=lightgreen, breakable}
\newtcolorbox{darkgreenbox}{colback=darkgreen!20!white, colframe=darkgreen, breakable}
\newtcolorbox{orangebox}{colback=lightorange!5!white, colframe=lightorange, breakable}

\usepackage{algorithm}
\usepackage{algorithmic}

\usepackage{newfloat}
\usepackage{listings}
\DeclareCaptionStyle{ruled}{labelfont=normalfont,labelsep=colon,strut=off} 
\floatstyle{ruled}
\newfloat{listing}{tb}{lst}{}
\floatname{listing}{Listing}
\usepackage{caption}
\usepackage{subcaption}

\title{AI Royalties \\ An IP Framework to Compensate Artists \& IP Holders for AI-Generated Content}
\author {
    Pablo Ducru \url{{p_ducru@mit.edu}}\textsuperscript{\rm 1,\rm 2, \rm 3}
    Jonathan Raiman \textsuperscript{\rm 4}
    Ronaldo Lemos \textsuperscript{\rm 2, 7}
    Clay Garner \textsuperscript{\rm 3, 9}\\
    George He \textsuperscript{\rm 5}
    Hanna Balcha \textsuperscript{\rm 6}
    Gabriel Souto \textsuperscript{\rm 3, 8}
    Sérgio Branco \textsuperscript{\rm 7}
    Celina Bottino \textsuperscript{\rm 7}
}
\affiliations {
    \textsuperscript{\rm 1}Massachusetts Institute of Technology, US\
    \textsuperscript{\rm 2}\'Ecole Polytechnique, France\
    \textsuperscript{\rm 3}Schwarzman Scholars, Tsinghua University, China\
    \textsuperscript{\rm 4}Universit\'e Paris-Saclay, France\
    \textsuperscript{\rm 5}Harvard Law School, US\
    \textsuperscript{\rm 6}NYU School of Law, US\
    \textsuperscript{\rm 7}Instituto de Tecnologia e Sociedade (ITS), Brazil\
    \textsuperscript{\rm 8} Laboratório de Políticas Públicas e Internet (LAPIN), Brazil\
    \textsuperscript{\rm 9}Stanford University, US\
    
}

\usepackage{bibentry}

\begin{document}
\maketitle

\begin{abstract}
This article investigates how AI-generated content can disrupt central revenue streams of the creative industries, in particular the collection of dividends from intellectual property (IP) rights. It reviews the IP and copyright questions related to the input and output of generative AI systems. A systematic method is proposed to assess whether AI-generated outputs, especially images, infringe previous copyrights, using a CLIP metric between images against historical copyright rulings. An examination (economic and technical feasibility) of previously proposed compensation frameworks reveals their financial implications for creatives and IP holders. Lastly, we propose a novel IP framework for compensation of artists and IP holders based on their published ``licensed AIs'' as a new medium and asset from which to collect AI royalties.
\end{abstract}

\section{Introduction \& Background}
 This decade has seen spectacular advances in generative artificial intelligence (AI) models -- notably generative adversarial networks (GANs)(2020), VQVAE (2021) and probabilistic diffusion models (DMs)(2015, 2022) -- ushering in a new area of high quality multimedia AI content generation, starting with image [DALL-E 2 \cite{ramesh2022hierarchical}, Imagen \cite{saharia2022photorealistic}, Parti \cite{yu2022scaling}, Stable Diffusion \cite{Rombach_2022_CVPR}, eDiff-I \cite{balaji2022ediffi}, \cite{ho2020denoising, karras2018progressive, rombach2022highresolution, dhariwal2021diffusion, saharia2021image}], and now pushing the frontier in AI video and sound generation [Imagen Video \cite{ho2022imagen}, Phenaki \cite{villegas2022phenaki}, Make-a-video \cite{singer2022make}, \cite{dhariwal_jukebox_2020, singer2022makeavideo}].

Increasing public access to these technologies has led to a boom in AI-generated content \cite{rombach2022textguided}, provoking a global debate around the legality of AI-generated content, the future of intellectual property rights (IPR), and potential disruption of the creative industries \cite{burk2020thirtysix, levin2023regulateai, compton2015casual, gordon2022co, shan2023glaze, lemley2020fair, levendowski2018copyright, grimmelmann2015copyright, sobel2017artificial, franceschelli2022copyright, eshraghian2020human, guadamuz2017androids, baio2022invasive, fjeld2017legal, margoni2022deeper, licencescomment2020, jointcomment2020, somepalli2022diffusion, heikkiläoptout2022, dotla, hertzmann2018can, henderson2023foundation, huang2023generative, elish2019moral, epstein2020gets, watson2019rhetoric, reeves1996media, ramalho2017}.

At the heart of concerns about IPR, legality, and AI-generated content lies the question of compensating creatives and IP-holders for their work.


Any such successful compensation framework will be based on contractual agreements, underpinned by some enforceable IP rights. Importantly, IPRs grant the power to exclude others from infringing on the IPR. This exclusivity (often limited in time) is at the heart of IP law and enables artists, creatives, and IP-holders to enforce and collect dividends from the revenues generated by their IPRs. These revenues are then contractually split amongst different stakeholders by a series of licensing frameworks. 
    
As of now, limited frameworks have been proposed to compensate IP-holders in a future where content is AI-generated -- most are based on paying to access the data necessary to train AI systems, and none have found widespread adoption from the creative industries \cite{grimes_twitter_2023, gapper2023, okeefe_windfall_2020}.
 
In this article, we focus on the key question of how to compensate creatives and IP-holders in a world of AI-generated content.

We first review the IP legal questions around input (training) and output (inference) of generative AI systems -- focusing on whether creatives and IP-holders can claim compensation for copyright infringement under US law. 
This prompts us to propose, focusing on the case of image generation, a systemic method to evaluate whether or not an AI-generated output violates previous copyright -- and assess the performance of this metric (CLIP) on a series of copyright rulings. 

Subsequently, we document previously proposed compensation frameworks, and estimate their monetary outcomes for artists and IP holders.

Finally, we propose a new IP framework based on officially ``licensed AIs'', viewed as a new format (and medium) for artistic creative expression, and a new asset from which creatives and IP-holders can collect AI royalties. We discuss the grounds (contractual and rights) for enforcing this new IP framework (extending copyright, trademark, and rights of publicity, to AI models).

\section{\label{sec: IP Problems of Generative AI: Training and Outputs} IP Problems of Generative AI: Training and Outputs}
U.S. law recognizes several classes of IPR that protect creative expression. These include:
    \begin{itemize}
        \item \textit{Copyright}: Protects original works of authorship, including literary, dramatic, musical, and artistic works \cite{uscopyrightoffice2022}.
        \item \textit{Trademark}: Protects brand names, logos, and other identifiers of source \cite{uspto2013lanham}.
        \item \textit{Rights of Publicity}: Protects an individual's right to exercise control over the commercial exploitation of their name, image, likeness, or other distinctive personal characteristics (persona) \cite{nimmer1954}. These rights prevent unauthorized use of an individual's identity for commercial purposes without obtaining their explicit consent. Unlike the Copyright and Trademark, there is no  federal law that explicitly safeguards an individual's right of publicity. Instead, the protection of this right is granted and regulated primarily by state laws through common law, statute, or both \cite{californiaCivilCode3344a, nycivilrights50F}.
    \end{itemize}

In the U.S., several tort claims arise from violations of copyright, trademark, and rights of publicity. These include: \textit{Infringement}: This pertains to unauthorized IP usage, governed by copyright, trademark, or rights of publicity laws \cite{rozansky2021protecting, uspto2023trademarkinfringement, copyrightgov2023infringement}; \textit{Dilution} (diminishing trademark \cite{hr1295_1995}); \textit{False Advertising} (misleading promotion \cite{uscode2023trademarks}); \textit{Misappropriation} (unauthorized use of valuable asset such as a trade secret \cite{utsa_1985}); \textit{Unfair Competition} (dishonest industry practice \cite{cwru2023intellectual, aba2006businesstorts}).

Specifically, copyright infringement is assessed based on four fair use factors. These are outlined in § 107 of Title 17, termed "Limitations on exclusive rights: Fair use" \cite{uscopyrightoffice_fairUse}:\\

\begin{greybox}
\textbf{17 U.S.C. §107: Four Factors for Copyright Fair Use}\label{box: Four factors for Copyright fair use}

\begin{enumerate}
    \item The purpose and character of the use, including whether such use is of a commercial nature or is for nonprofit educational purposes
    \item The nature of the copyrighted work
    \item The amount and substantiality of the portion used in relation to the copyrighted work as a whole
    \item The effect of the use upon the potential market for or value of the copyrighted work
\end{enumerate}
\end{greybox}

Artists and the wider creative industries earn revenue through the monetization of IPR -- including royalties from copyright (creative works) and licensing related to rights of publicity and trademarks \cite{wipo2014monetization, rothman2018right}.
Generative AI has prompted a backlash of copyright infringement lawsuits from the creative industries, for producing art-like output and training on copyrighted material scrapped from the Internet, with neither consent nor compensation \cite{henderson2023foundation, gettyvstability, coscarelli2023, andersen2023}. 
Other creative industry figures have sought to proactively integrate AI into their IP strategies, creating digital replicas of celebrity likenesses and corresponding licensing schemes \cite{coffee2023celebrities, grimes_twitter_2023}. 
Academic and industry discussion is centered on two areas: foundation model training -- the "input" -- and whether it constitutes fair use or copyright infringement,  \cite{henderson2023foundation, epstein2023art, lemley2021fair, bonadio2020artificial, hsu2023can, metz2022lawsuit, lemley2020fair, sobel2017artificial, levendowski2018copyright, Williams2023, samuelson2023generative}, and to what extent (and under which circumstances) the outputs of AI systems can be copyrighted \cite{samuelson2023generative}.   
Varying regulatory perspectives have emerged on whether AI training falls under fair use \cite{kii2023reflecting, band_israel_2023} and whether AI output is afforded intellectual property rights, including works of mixed human and AI authorship \cite{copyright_office_2023, thaler2021, thaler2021sa}.

Concerning our key question of compensation for AI-generated content, if training an AI system on data constitutes a violation of IP rights, then the IP-holders can claim compensation -- this is the approach taken by the recent wave of lawsuits against the use of the LAION dataset and open image-generating AI systems (such as  Microsoft \& OpenAI \cite{metz2022lawsuit}, Meta \cite{meta_complaint}, Stability.ai \cite{gettyvstability}, Midjourney \cite{stability_ai_complaint}).
Some proposed compensation schemes studied in section \ref{Previous compensation frameworks} rely on compensation for training, somewhat resting on the assumption that training is not fair use.  
New registration guidance released in March 2023 reflects the U.S. Copyright Office's increasingly nuanced stance towards AI authorship -- suggesting that there will be more paths toward extending some IPR to AI output \cite{copyright_office_2023}.

Two key unanswered questions emerge from this discussion:
        \begin{enumerate}
            \item Does an AI-generated output infringe someone’s IPRs (in particular copyright)?
            \item If the output does not infringe IPRs, what framework could compensate IP-holders for inspiring AI-generated content, and on what grounds could it be enforced? (we propose AI publishers and AI royalties in section \ref{A Proposal for AI Royalties}).
        \end{enumerate}


We propose a systematic AI method to evaluate whether an AI-generated output infringes previous IP rights, and assess its performance on a series of rulings. 

For standardization, we focus on image copyright infringement, using an approach that is readily extendable (with corresponding metrics) to other forms of IPRs (trademark, image, name, likeness, etc.) and creative expression protected by copyright (sound, video, 3D graphics and games, text, etc.). 
Of the four factors considered when ruling for copyright infringement \ref{box: Four factors for Copyright fair use}, we focus on a metric for sufficient transformation the image itself (factor 3). For a broader analysis encompassing all four factors, this metric could be combined in a multi-modal AI system incorporating predictive treatments of court rulings, such as previous specifically developed LLMs \cite{li_court_2021}.

\begin{figure}
\centering
\begin{subfigure}{\columnwidth}
    \label{fig:first}
    \includegraphics[width=0.49\columnwidth]{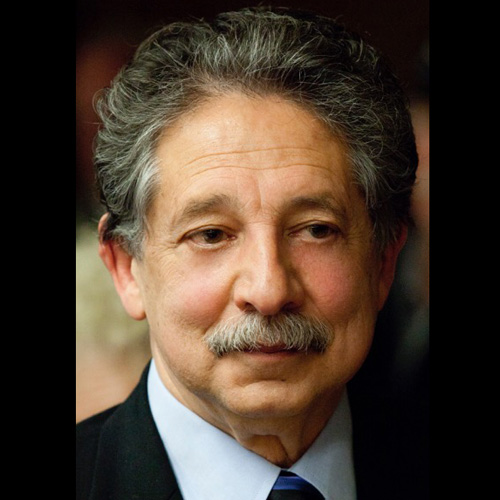}
    \includegraphics[width=0.49\columnwidth]{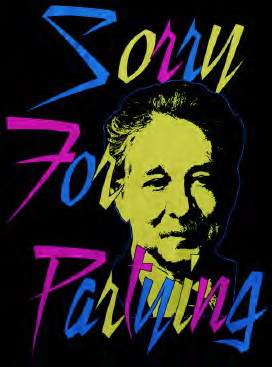}
    \caption{Kienitz v. Sconnie Nation: fair use, CLIP metric = 0.479}
    \label{fig:Kienitz v. Sconnie Nation}
\end{subfigure}
\hfill
\begin{subfigure}{\columnwidth}
    \includegraphics[width=0.49\columnwidth]{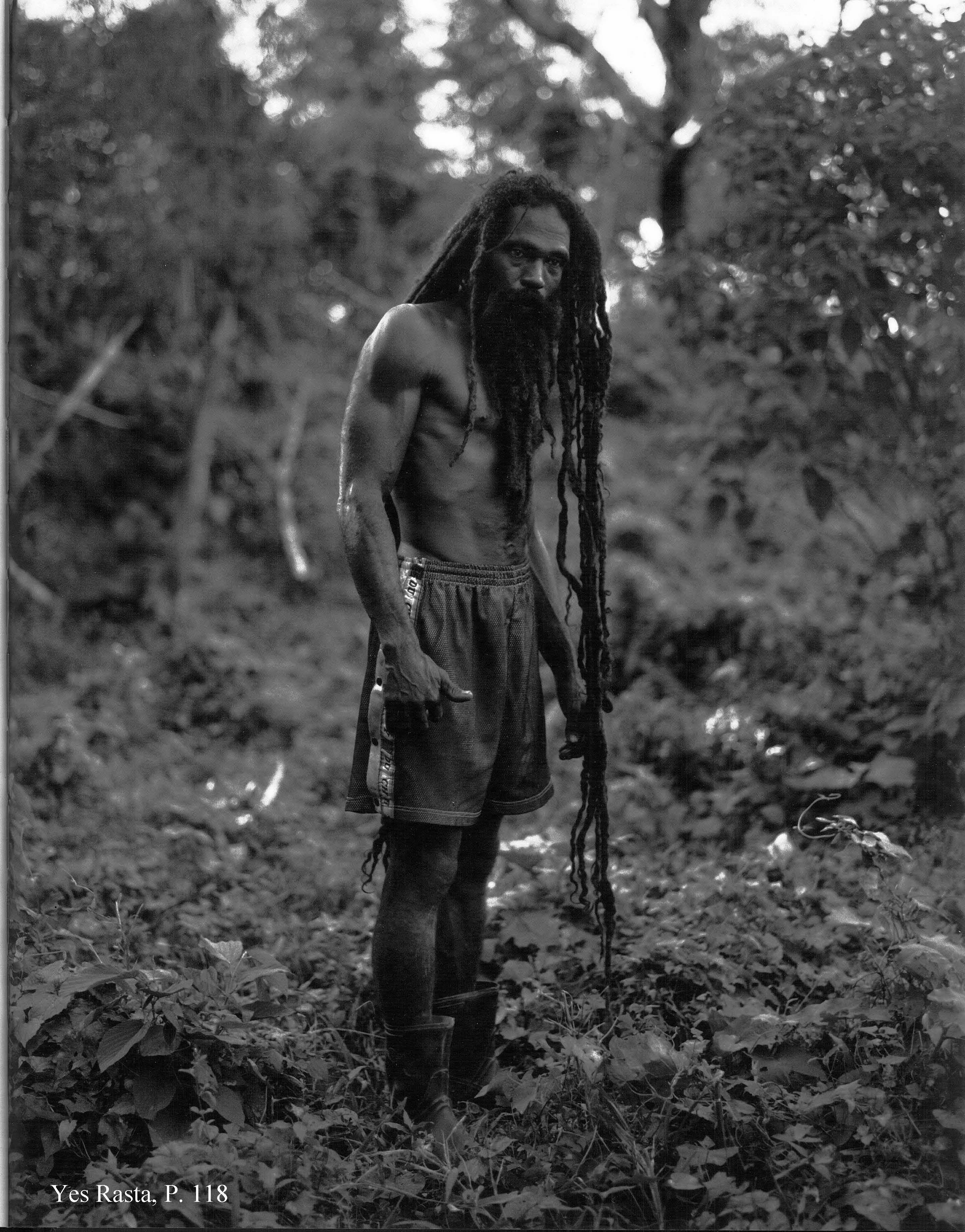}
    \includegraphics[width=0.49\columnwidth]{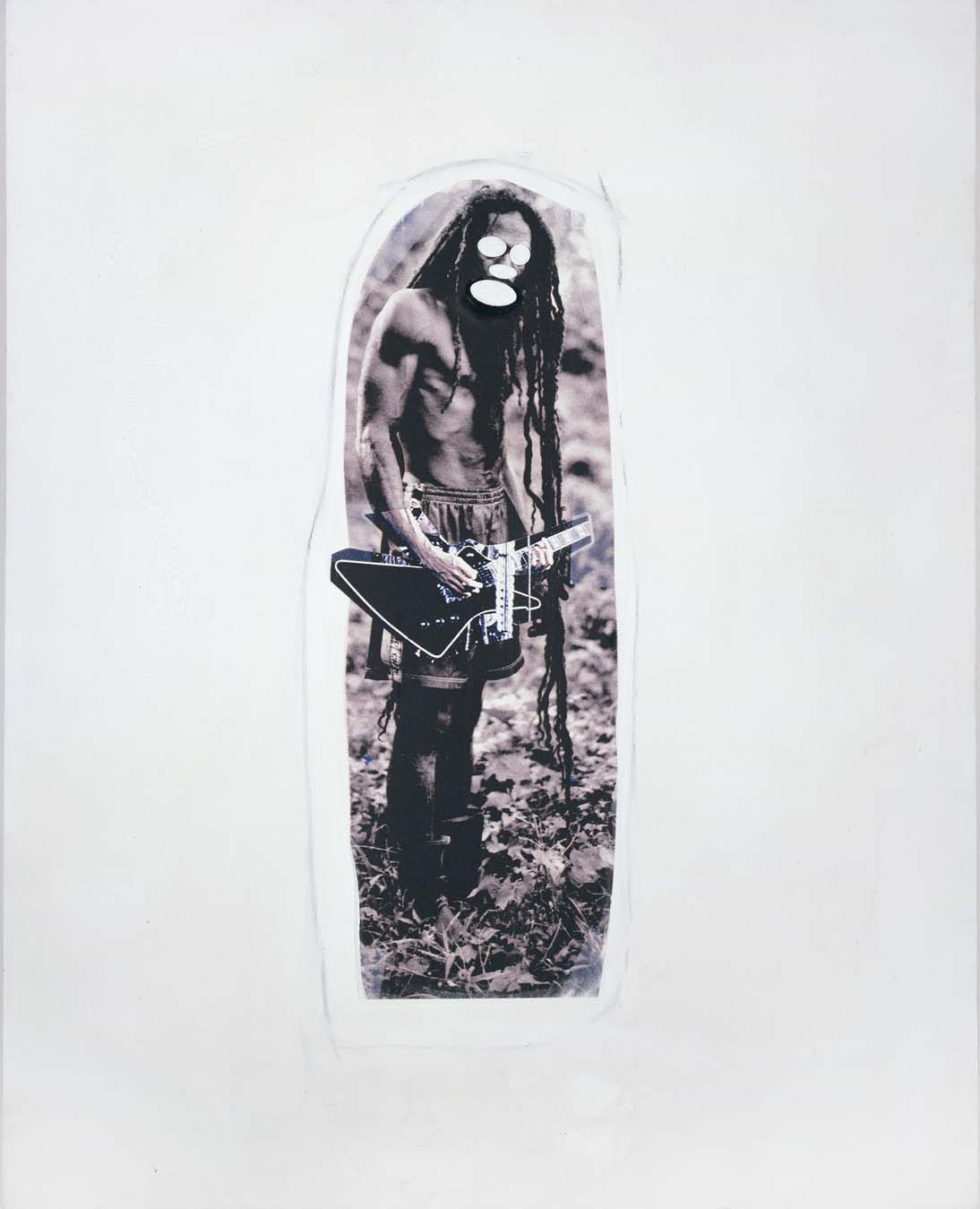}
    \caption{Cariou v. Prince: probably not fair use, CLIP metric = 0.776}
    \label{fig:Cariou v. Prince}
\end{subfigure}
\hfill
\begin{subfigure}{\columnwidth}
    \includegraphics[width=0.49\columnwidth]{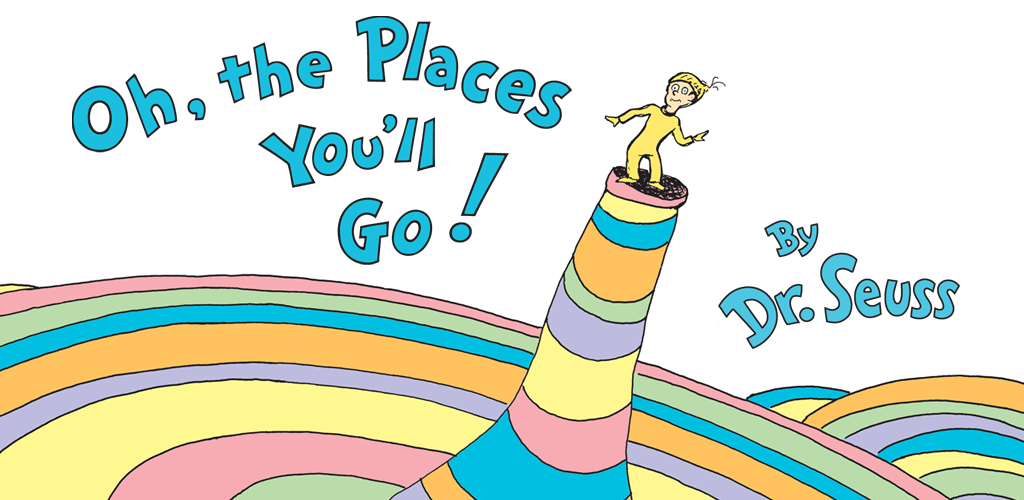}
    \includegraphics[width=0.49\columnwidth]{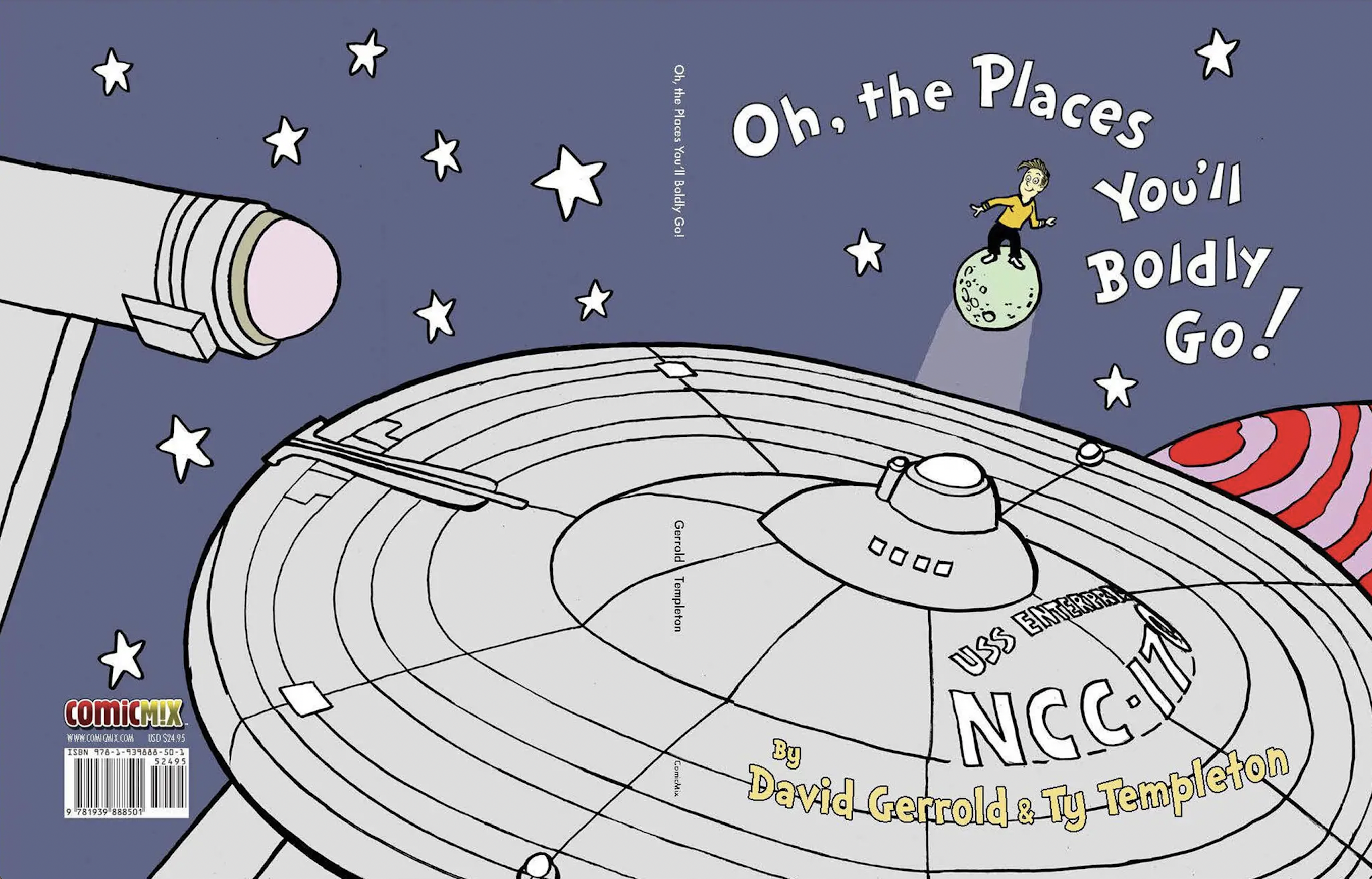}
    \caption{Dr. Seuss v. ComicMix: not fair use (copyright infringement), CLIP metric = 0.723}
    \label{fig:Warhol v. Goldsmith}
\end{subfigure}
\begin{subfigure}{\columnwidth}
    \includegraphics[width=0.49\columnwidth]{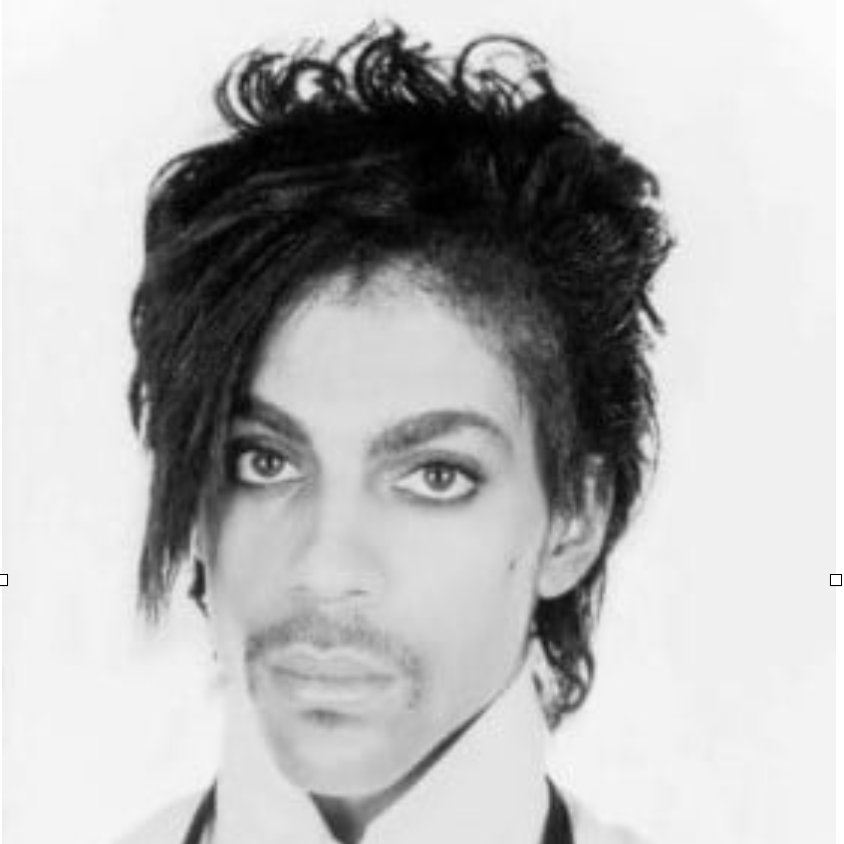}
    \includegraphics[width=0.49\columnwidth]{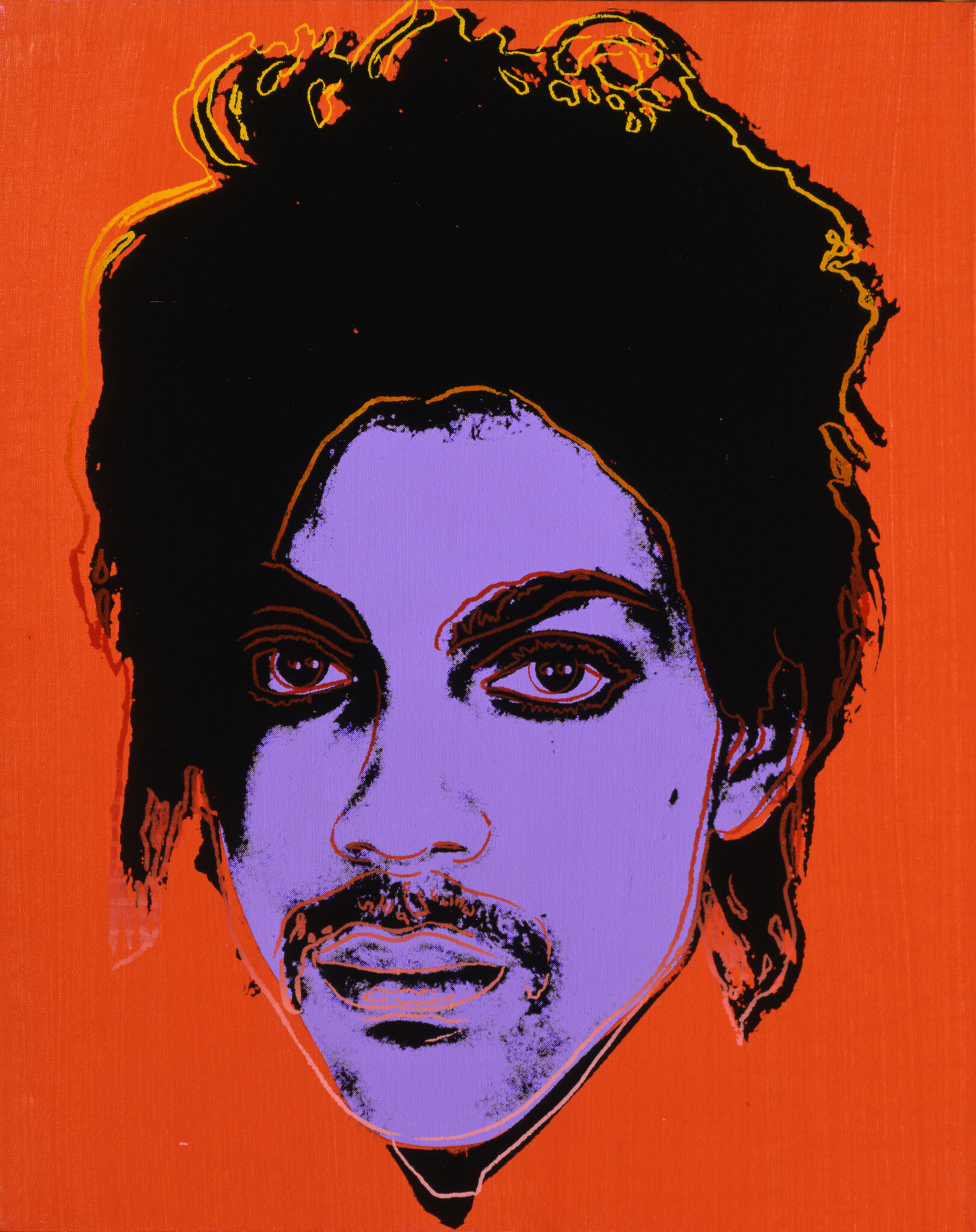}
    \caption{Warhol foundation v. Goldsmith: not fair use (copyright infringement), CLIP metric = 0.852}
    \label{fig:Warhol v. Goldsmith}
\end{subfigure}
\caption{Copyright cases and CLIP-based metric}
\label{fig:Copyright cases and metric}
\end{figure}

\begin{figure}[t]
\centering
\includegraphics[width=0.9\columnwidth]{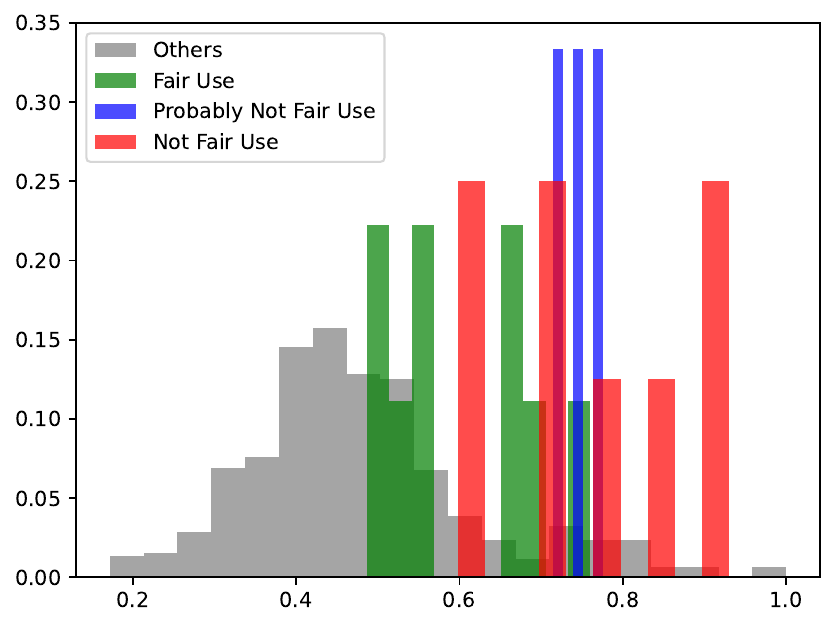}
\caption{CLIP Metric and Copyright Infringement}
\label{fig:CLIP for fair-use metric}
\end{figure}

\begin{orangebox}
\textbf{Copyright infringement metric}
\\
Metrics to measure whether or not a new image constitutes sufficient transformation for fair use or violates previous copyrights must: 
    \begin{itemize}
        \item Identify identical images beyond formats, aspect ratios, or indistinguishable pixel alterations
        \item Identify transformations of the same image
        \item Identify different images referencing identical objects or subjects 
        \item Integrate the subjective elements of precedent jury's decisions and court rulings
    \end{itemize}

Since these combine technical and subjective social aspects, such metric should extend beyond mere visual image analysis.
We propose to use a recent AI metric that combines pixel processing of images with contextual language, called CLIP. 
CLIP models \cite{radford2021learning} were made freely available by research organizations such as OpenAI, Hugging Face, Stability.AI, etc. 
The family of CLIP image understanding models is trained to recognize the similarity between an image and a caption. They thus combine elements of visual and language intelligence of images that could be repurposed to assess the level of transformation an original image has undergone, and interpret it in light of precedent copyright rulings. \\

\textbf{Dataset of U.S. Copyright Images Rulings:} We constituted a dataset of 10 copyright fair use rulings (accessible here \footnote{\url{https://github.com/Pablo-Ducru/ai-royalties}}), totaling 20 pairs of contested works, opposing 14 original works to derivative ones, where the ruling was primarily decided along the criteria of sufficient transformation for fair use or not (factor \#3 -- ``the amount and substantiality of the portion used in relation to the copyrighted work as a whole.''). There is no straightforward way to do this: there are relatively few cases limited to factor \#3 and visual image; and some of the cases (such as Andy Warhol Foundation for the Visual Arts, Inc. v. Goldsmith) were also ruled with other factors contributing significantly. It is likely many cases were settled out of court or not contested, and these cases are not represented in the dataset.
\\

\textbf{Analysis:} We measured the CLIP distance for every pair, documenting the results in figure \ref{fig:CLIP for fair-use metric} and table \ref{Tab: CLIP distance between cases}.
In figure \ref{fig:CLIP for fair-use metric}, the CLIP distance between images ruled as fair use is documented in green, that between those ruled not fair use (copyright infringements) in red, those in blue were deemed probably not fair use, while the CLIP distance between all other pairs is documented in grey. 

Though we are in a heavily undersampled regime with low statistical power (only 20 pairs of copyright infringement image cases) we can observe a distinction between uncontested images, and contested ones, with the average CLIP distance between uncontested image pairs around 0.5, while that for those contested around 0.69. 
Moreover, for contested images, CLIP seems to be able to discern between those ruled fair use (mean of 0.6), and those infringing copyright (mean of 0.76), with a resolution beyond the first standard deviation.
Examples of contested image pairs are documented in figure \ref{fig:Copyright cases and metric}, along with their respective CLIP distance. Though more cases would be necessary to improve the statistical power and resolution, this approach shows promise towards building an AI system to detect copyright infringement. Note that starting from a CLIP based metric with the thresholds proposed in table \ref{Tab: Copyright and fair use metric (CLIP-based)}, additional training on an embedding specialized on copyright cases is likely to produce a metric with a more statistical resolution power.
\end{orangebox}

\begin{table}[t]
\caption{CLIP distance between contested cases}
\label{Tab: CLIP distance between cases}
\centering
\begin{tabular}{lcr}
\hline
$\mathrm{CLIP}$ Distance  &  Mean &  Standard Deviation \\
\hline
\hspace{1mm} $\cdot$ Fair Use & $0.604$  & $ \pm 0.093$ \\ 
\hspace{1mm} $\cdot$ Not Fair Use &  $0.764 $ & $ \pm 0.123  $ \\
\hline
\end{tabular}
\end{table}

\begin{table}[t]
\caption{CLIP-Based Thresholds for Copyright \& Fair Use}
\label{Tab: Copyright and fair use metric (CLIP-based)}
\centering
\begin{tabular}{lr}
\hline
If Contested & $ \mathrm{CLIP}$ Distance \\
\hline
Copyright safe & $ \mathrm{CLIP} \leq 0.6 $ \\
Likely fair use  & $ 0.6 < \mathrm{CLIP} \leq 0.7 $ \\
Likely copyright infringement &  $ 0.7 < \mathrm{CLIP}  $ \\
\hline
\end{tabular}
\end{table}


\section{\label{Previous compensation frameworks}Previously Proposed Compensation Frameworks}
As of now, three main frameworks have been proposed to compensate IP-holders for AI-generated content:
\begin{enumerate}
    \item Universal income from Windfall AI profits (section \ref{sec: No Direct Compensation - No Shared Ownership - Windfall Clause}).
    \item Pay to train, proportionally to contribution to training dataset (section \ref{sec: Compensate to train - Pay IP holders for contribution to training datasets}).
    \item Pay to train \& inspire, proportionally to ``what input training data inspired this output?'' (section \ref{sec: Compensate to train and inspire - Trace back to the data that inspired the output}). 
\end{enumerate}
The latter two implicitly rest on assuming training is not fair use, and the last one seeks to make a direct connection between inputs and outputs of an AI system (at inference). We assess compensation for these frameworks across examples from three categories of artists, looking at relative output volume and fame: the median Shutterstock stock photo contributor (2,000 low-fame images) \cite{kneschke2023shutterstock}, the emerging artist Sam Yang (500 medium-fame images), and the renowned artist Claude Monet (2,000 high-fame images).

\subsection{\label{sec: No Direct Compensation - No Shared Ownership - Windfall Clause}No Direct Compensation -- Windfall Clause}

 \textbf{No shared ownership} is a scenario where both training is fair use (does not infringe IPRs), and the outputs of generative AI systems cannot be legally owned through copyright or another IPR mechanism. 
 In this case, creatives and IP-holders cannot be compensated for contributing to the training of an AI system, nor can they own the AI outputs. If AI-generated content ends up being very valuable, the artists are simply displaced. 
 If artists and IP holders are not compensated for inspiring, contributing to directly, or prompting AI art, this raises serious ethical, moral, and financial concerns about the viability of creative work \cite{epstein2023art}.
    
Anticipating a world where AI-caused displacement negatively impacts labor participation, a proposal has emerged for a ``windfall clause'' as an ``\textit{ex ante} commitment by AI firms to donate a significant amount of any eventual extremely large profit'' \cite{okeefe_windfall_2020}. The authors propose that this commitment scales ``based on signatory's profits as [a] portion of gross world product'' -- for example, with a marginal clause obligation (portion of marginal profits) of 1\% to be donated for profits 0.1\%-1\% of gross world product  \cite{okeefe_windfall_2020}. 
We here estimate the compensation this Windfall Clause would yield for displaced artists.

\begin{greenbox}
\textbf{Windfall Clause, Adapted for US}:
\begin{darkgreenbox}
The proposed windfall clause could offer an ``unemployability insurance'' to someone who loses their job due to AI. \cite{okeefe_windfall_2020}.
\end{darkgreenbox}
In 2030, assume the U.S. workforce is 165.4 million \cite{BLS2021}, 30\% of US labor automated \cite{ellingrud2023generative} and AI developers' profits equal 0.5\% of US GDP (\$35 trillion). \\
\textbf{Compensation:} With AI profits at \$175 billion, a 1\% Windfall Clause would generate \$1.75 billion for the estimated 50 million displaced workers (30\% automation), amounting to \$35 per year per displaced worker. Profits of 5\% of GDP would lead to a 20\% clause, yielding \$700 per year per displaced worker.
\end{greenbox}

\subsection{\label{sec: Compensate to train - Pay IP holders for contribution to training datasets} Compensate to train - Pay IP holders for contribution to training datasets} 

\textbf{Pay-to-train} is a proposed compensation scheme consisting of paying IP holders for the use of their IP in training datasets of generative AI models \cite{gordonlevitt2023ai, miller2023adobe, murphy2023pay}. Compensation for and -- by extension -- consent to train are at the core of the lawsuits brought against AI developers, including Microsoft \& OpenAI \cite{metz2022lawsuit}, Meta \cite{meta_complaint}, Stability.ai \cite{gettyvstability}, Midjourney \cite{stability_ai_complaint}, etc. Regardless of whether or not training is considered fair use \cite{henderson2023foundation}, current pay-to-train compensations are based on contributions as a percentage of the total dataset -- often hundreds of millions of unique works.


\begin{greenbox}
\textbf{Compensate to Train (Shutterstock's AI Data Contributor Fund)}:
Shutterstock constituted a large training dataset (734 million images by Q3 2023), licensing it to earn revenues, and contributors earn a share of this revenue on a pro-rata basis of their content volume in the dataset \cite{Shutterstock2023}.
\begin{darkgreenbox}
Contributors to Shutterstock's AI models will get a share of the contract value based on the amount of their content in the licensed datasets \cite{Shutterstock2023}.
\end{darkgreenbox}

\textbf{Compensation}: A recent survey of revenue to contributors from AI generated content \cite{kneschke2023shutterstock} found the average contributor has 6,343 images, with a median of 2,112 images, and estimated a total payout to contributors of \$4.4 million in the first 6 months. Assuming Shutterstock's first 6 months revenue is \$430M, this means about $1\%$ of total revenues went to the contributor fund. This yielded average revenues per image of \$0.008, and (median of \$0.007), and corresponding 6-month payout averages of \$46 (\$92/yr), with a median of \$18.5 (\$37 /yr) per contributor -- for a total dataset with 615 million images \cite{kneschke2023shutterstock}.
In comparison, traditional compensation for contributors to stock imaging have averaged \$0.24 per image over a year (though royalty-free sales can generate as much as \$100 \cite{broz2023stockphotos}). This means a median contributor with 2,112 images can expect \$507/yr (and the average one with 6,343 some \$1,522/yr).

Let $d_c$ be the percentage of total revenues for AI-generated outputs that are distributed to the contributors of the AI dataset. 
Without knowing the fraction of total revenues that was generated by AI contracts, we cannot provide a clear estimate of $d_c$ for today's Shuttershock Contributor Fund.
However, we could consider $d_c = 55\%$, since YouTube and Instagram share 55\% of ad revenue with creators, which has been described as ``industry standard'' \cite{dayal2022youtube}.
If we assume that, by the end of 2024, 90\% of Shutterstock images will be AI-generated, and that the total revenue is \$1 billion/yr, with the dataset size growing to 1 billion images. A median contributor with 2,112 images in this scenario would generate \$1,045/yr (and the average revenues from 6,343 contributions would yield \$3,140/yr).

One should consider the different dynamics at play between stock media (often priced in volume) and famous artwork (often priced on celebrity). For instance, considering Monet and Van Gogh, who each produced around 2,000 works in their lifetimes, a contributor fund model based on pro-rata contributions to the above dataset (by volume) would yield \$990/yr for each artist. Moreover, a different industry standard for AI revenue splitting may emerge than that of the ads model -- especially if IPR are not infringed or if enforcement is intractable. If, for example, our simulation were $d_c =10\% $ distributed instead of $d_c =55\% $, Monet would make \$180/yr.

\end{greenbox}

\subsection{\label{sec: Compensate to train and inspire - Trace back to the data that inspired the output}Compensate to train \& inspire -- Trace back to the data that inspired the output}

 \textbf{Pay to train \& inspire} is a compensation scheme proposed as a potential solution to the volume (stock media) versus value/recognition (say Van Gogh or Monet) problem of a flat pro-rata pay-to-train scheme (note that both rest on training not being fair use).
 Some creatives and IP-holders have expressed the wish to be compensated proportionately to how much the content their contributed to the training dataset is ``being consumed’’ by the AI model and its users, that is according to ``how much was this output influenced by my data’’. 
 Conceptually this would mean calculating an ``influence function’’ for each training datapoint, given an output and prompt.
 Philosophically it is akin to asking someone to trace back, from everything they have learned, how much each thing influenced what they just wrote. 

Let $X = \left\{ x_i \right\}$ be a training dataset on which generative AI model $g_X$ is trained. The model is prompted with inputs and conditions $c$, yielding output $y = g_X(c) $. Consider one could build an ``influence function’’, $f(x;y,c) \in [0,1]$ which estimates the percentage to which each datapoint $x_i$ contributed to generating output $y$, such that:
\begin{equation}
     \sum_{x_i \in X } f(x_i;y,c) = 1
     \label{eq: influence function normalization}
\end{equation}

\begin{orangebox}
\textbf{Tracing back what training data influenced the AI-generated output}\\
A first way to compute influence functions from training data was introduced in \cite{koh2017understanding}.
In theory, one could compute these and normalize them to calculate $f(x;y,c)$ that satisfy (\ref{eq: influence function normalization}). 
However, this is intractable in practice because the influence functions need to be computed for every training point (with respect to an output), including a Hessian of the size of the parameter space. Though various methods have made this computation more efficient \cite{koh2017understanding}, their computational complexity is still prohibitive for large AI systems with tens of billions of parameters, trained on billions of datapoints. 

Since, discussions about traceability, explainability, and fair-use of training, have proposed various methods to trace back the data that influenced a given output \cite{henderson2023foundation, wang2023evaluating, somepalli2022diffusion, koh2017understanding}.
In particular, a recent one based on model and instance attribution \cite{wang2023evaluating} is promising for the creative industries. 
It proposes to create various fine-tuned models (from a foundation base), which can be attributed to a given dataset (say specific to an artist or IP-holder) -- each for a specific concept (object, likeness, style). 
For a given output, the method performs model attribution -- which models generated this output? -- and within the model, extracts soft influence scores to determine which images influenced the output \cite{wang2023evaluating}. 

Though promising, these methods remain costly and more work is needed before industry adoption. For instance, they only consider the training data. Yet the prompt and constraints $c$ (specially if an image) are also instrumental to the output — how to account for these? Parsing out the influence of training, fine-tuning, and prompting (in particular augmented retrieval), is an open problem which is likely to only have somewhat subjective answers (highly dependent on modelization and implementation).

\end{orangebox}


\begin{greenbox}
\textbf{Compensate to train \& inspire}\\
In pay-to-train-\&-influence on dataset $X$, a contributor $A$ of data $X_A \in X$ to would be compensated for output $y$ according to $\sum_{x_i \in X_A } f(x_i;y,c)$. In a revenue-share scheme where output $y$ generates revenues $r(y)$, the compensation $C_A$ to IP-holder $A$ for all outputs $Y$ generated over a given time period would be proportional to:
\begin{equation}
     C_A \propto \sum_{  y \in Y  }  r(y)  \sum_{ x_i \in X_A } f(x_i;y,c)
     \label{eq: proportional compensation}
\end{equation}
\end{greenbox}

In practice, estimating such influence functions $f(x_i;y,c)$ is both technically challenging (as reviewed), and may not really solve the compensation problem:
\begin{itemize}
	\item First, it does not necessarily solve the compensation problem, because it is possible to train an AI model without using any data of a given artist, and still produce outputs so close to the artists work as to infringe their IP.
\end{itemize}
\begin{orangebox}
\textbf{No need to train on your data to reproduce you}\\
 A generative AI model does not need to train on an artist's works to produce an output that closely mimics their style, or infringe on their IP. 
 For instance, one could train an AI model that produces highly similar works to Andy Warhol (to the point of possible infringement) -- without that model ever ``seeing'' any Andy Warhol works in the training dataset. Under the ``pay-to-train-\&-inspire'' logic, the inspiring artist would not get compensated. Such reproduction or inspiration without using the subject in the training data has long existed, for instance with likeness and eigenfaces \cite{turk1991eigenfaces}, and is the subject of recent developments such as ``Celeb Basis’’. \cite{yuan2023inserting}.
\end{orangebox}

\begin{itemize}
	\item Second, as AI-generated data (influenced by training data) is used to train new models (and therefore influence new AI-generated outputs), it opens a somewhat intractable problem of recursive fractionalization. 
     \item Third, it does not solve the ownership problem -- specially if training is fair use. Even if clear attribution were drawn to original IP, it would not yield ownership rights to the outputs. For instance, style is not protected -- a work can be ``heavily inspired by Andy Warhol'' and, with sufficient transformation, not infringe. As such, a contractual pay-to-train-\&-influence compensation would be a departure from current creative industry practices.
\end{itemize}

\section{\label{AI Royalties}AI Royalties -- Artists partner with AI companies and share revenues}


\begin{table*}[t]
\caption{Compensation schemes comparison}
\label{Tab: Compensation schemes comparison}
\centering
\begin{tabular}{l | l | l | l | l}
\hline
 & \begin{tabular}{@{}l@{}}No contributor \\ person\end{tabular} & \begin{tabular}{@{}l@{}}Stock media (median) \\ contributor\end{tabular} & \begin{tabular}{@{}l@{}}Artist (median) \\ Greg Rutkowski\end{tabular} & \begin{tabular}{@{}l@{}}Artist (famous) \\ Claude Monet\end{tabular} \\
\hline
\hline
Volume of works   &  0 & ~2000 & ~200 & ~2000 \\
\hline
Windfall  &  \$35/yr & \$35/yr & \$35/yr  & \$35/yr \\
Compensate-to-train  &  0 & \$1,000/yr & ~\$100/yr & ~\$1,000/yr \\
   \begin{tabular}{@{}l@{}} AI royalties (fame) \\  (Monet 1000x Rutkowski) \end{tabular} &  0 &  \$500/yr  & ~\$550/yr &  \$50,500/yr  \\
\begin{tabular}{@{}l@{}} AI royalties (fame) \\  (Rutkowski 1000x Monet) \end{tabular}   &  0 &  \$500/yr  & ~\$50,500/yr &  \$550/yr  \\
\hline
\end{tabular}
\end{table*}

We hereby propose a partnership framework to compensate creatives and IP-holders for AI-generated content based on the market usage and value of their AIs, rather than the volume they contributed (c.f. sections \ref{sec: Compensate to train - Pay IP holders for contribution to training datasets} \& \ref{sec: Compensate to train and inspire - Trace back to the data that inspired the output}). 

 \subsection{\label{sec: Contractual compensation: AI royalties} Contractual compensation: AI royalties}
 
We propose a collaborative arrangement that would treat holders of IPR as partners in the success of an AI model that would be built by an AI company in collaboration with the IP partner, and specifically dedicated to the named rightsholder’s IP. There is no need to revolutionize the existing legal framework for this to work. The IP partner and the AI company would enter into a contractual arrangement that accepts a priori the existing IPR of the rightsholder under the current legal framework with respect to outputs of the AI system. This proposal would recognize the exclusive rights of an IPR holder to the commercial exploitation of outputs that are substantially similar under copyright, derivative works under copyright, confusingly similar trademark uses or are readily identifiable uses of the holder’s rights of publicity.
As a partner, the named rightsholder would then participate in all of the revenues generated by the dedicated AI model.
The rightsholder would grant the AI company the right under their IPR to use the AI system to create outputs. In exchange, as a commercial partner, the rightsholder would receive a share of all of the revenue generated by the dedicated AI system.
Because the rightsholder will participate in all of the revenue generated by the dedicated AI system without regard to whether the revenue is derived from outputs that would otherwise infringe the rightsholder’s IPR, there is no need to make fine legal distinctions with the named rightsholder partner on an output-by-output basis.
As a partner, the rightsholder could participate also in establishing acceptable use and editorial guidelines with respect to the output, thereby allowing the rightsholder to oversee and protect enduring brand and name value.  
The foregoing collaborative partnership reduces many points of potential friction by aligning the interests of the parties.

To illustrate this framework, we compare the results from the compensate-to-train model discussed above, to this IP rightsholder partnership with an AI company to build and commercialize their AI. We consider that overall, 50\% of revenues will go to stock media having trained the base models, and 50\% will go to AI models dedicated to specific rightsholders' IP, taking the example of two artists, Claude Monet, who produced about 2,000 painting throughout his life, and Greg Rutkowski, who has produced about 200 to date. Claude Monet being one of the most famous impressionist painters, and given the power-law distributions of fame \cite{simkin_mathematical_2013, ramirez_quantitative_2018, prinz_concentration_2022, gustar_fame_2019, gustar_laws_2020, etro_power-laws_2018, visser_zipfs_2013} we can a priori estimate that 1000 times more people know of Monet than Rutkowski. We compare in table \ref{Tab: Compensation schemes comparison} the compensation the different schemes could amount to, in orders of magnitude. The case of Greg Rutkowski is particularly interesting here because this artist became very popular in prompting for AI-generated imaged \cite{Greg_Rutkowski_MIT_tech_review_2022, Greg_Rutkowski_Opt-out-LoRA-in_2023}. This means that with the type of scheme that we propose, where a specific Rutkowski AI model could be attributable and licensed, Rutkowski could have collected significant revenues from the heavy usage people did of this AI style (as we suggest with the last row of table \ref{Tab: Compensation schemes comparison}).

\subsection{\label{sec: Enforcement and defensibly: legal, contractual, moral, commercial} Enforcement and defensibly}

This framework could be enforced on two grounds: 

\textbf{Legal \& contractual:} Legally, copyright, trademark, and rights of publicity would underpin the partnership. Contractually, licensing rights and terms of service could restrict usage (e.g. personal or commercial use). Misappropriation or other torts could also be claimed for undermining the livelihood of an artist because a ``non-licensed AI'' is provoking a loss of revenue in a newly established AI market. This could make paying AI royalties to artists and IP-holders a \textit{de facto} industry \textit{modus operandis}. 

 \textbf{Technology:} AI-generated content IDs with tracking systems, as well as automatic contracts could help enforcement -- e.g. takedowns from automatic identification of copyright infringement (YouTube), or content moderation on social platforms. Also, it may be impossible to technologically suppress online piracy, but convenient technology can funnel the vast majority of users to a legal paid service -- c.f. streaming content platforms (Spotify, Netflix, Disney+, etc.).

\subsection{\label{Forward Looking Agenda}Forward Looking Agenda}
In examining the future of AI-generated content, we should recall the primary objective of intellectual property (IP) protection: to support and incentivize human creativity by offering financial benefits to creators.
As AI starts to create content of its own, it will be necessary to determine clear authorship rights and adapt IP frameworks that emphasize human contribution. Additionally, we should consider the potential for value dilution from AI-generated content and find solutions that address this issue without hindering creative reinterpretation.

\bibliography{ai_royalties_for_multimedia_generative_ai_models}

\begin{thebibliography}{105}
\providecommand{\natexlab}[1]{#1}

\bibitem[{{15 U.S.C. § 1125}(2021)}]{uscode2023trademarks}
{15 U.S.C. § 1125}. 2021.
\newblock 15 USC 1125: False designations of origin, false descriptions, and dilution forbidden.
\newblock U.S. Code.
\newblock Subchapter III - General Provisions.

\bibitem[{{ABA Section of Antitrust Law}(2006)}]{aba2006businesstorts}
{ABA Section of Antitrust Law}. 2006.
\newblock \emph{Business Torts and Unfair Competition Handbook}.
\newblock American Bar Association.
\newblock Federal Law of Unfair Competition.

\bibitem[{{Andersen et al v. Stability AI}(2023)}]{andersen2023}
{Andersen et al v. Stability AI}. 2023.
\newblock Andersen et al v. Stability AI Ltd. et al, Docket No. 3:23-cv-00201.
\newblock N.D. Cal.

\bibitem[{Baio(2022)}]{baio2022invasive}
Baio, A. 2022.
\newblock Invasive Diffusion: How one unwilling illustrator found herself turned into an AI model.
\newblock \emph{Waxy.org}.

\bibitem[{Balaji et~al.(2022)Balaji, Nah, Huang, Vahdat, Song, Kreis, Aittala, Aila, Laine, Catanzaro et~al.}]{balaji2022ediffi}
Balaji, Y.; Nah, S.; Huang, X.; Vahdat, A.; Song, J.; Kreis, K.; Aittala, M.; Aila, T.; Laine, S.; Catanzaro, B.; et~al. 2022.
\newblock eDiffi: Text-to-Image Diffusion Models with an Ensemble of Expert Denoisers.
\newblock \emph{arXiv preprint arXiv:2211.01324}.

\bibitem[{Band(2023)}]{band_israel_2023}
Band, J. 2023.
\newblock Israel Ministry of Justice Issues Opinion Supporting the Use of Copyrighted Works for Machine Learning.
\newblock \url{https://www.project-disco.org/intellectual-property/011823-israel-ministry-of-justice-issues-opinion-supporting-the-use-of-copyrighted-works-for-machine-learning/}.

\bibitem[{Bonadio and McDonagh(2020)}]{bonadio2020artificial}
Bonadio, E.; and McDonagh, L. 2020.
\newblock Artificial intelligence as producer and consumer of copyright works: Evaluating the consequences of algorithmic creativity.
\newblock \emph{Intellectual Property Quarterly}, 2: 112--137.

\bibitem[{Broz(2023)}]{broz2023stockphotos}
Broz, M. 2023.
\newblock How Much Can You Make Selling Stock Photos? (2023 Data).
\newblock Accessed on 2023-08-15.

\bibitem[{Burk(2020)}]{burk2020thirtysix}
Burk, D.~L. 2020.
\newblock Thirty-Six Views of Copyright Authorship, By Jackson Pollock.
\newblock \emph{Houston Law Review}, 58.
\newblock UC Irvine School of Law Research Paper No. 2020-40.

\bibitem[{{California Civil Code § 3344(a)}(1971)}]{californiaCivilCode3344a}
{California Civil Code § 3344(a)}. 1971.
\newblock California Civil Code § 3344(a) - Use of another's name, voice, signature, photograph, or likeness in advertising or soliciting without prior consent.

\bibitem[{{Case Western Reserve School of Law}(2023)}]{cwru2023intellectual}
{Case Western Reserve School of Law}. 2023.
\newblock Intellectual Property Law.
\newblock This guide covers basic resources for researching intellectual property (IP) law: copyright, patents, trademarks, and more.

\bibitem[{{Civil Rights (CVR) Chapter 6, Article 5 - § 50-f. Right of publicity}(2020)}]{nycivilrights50F}
{Civil Rights (CVR) Chapter 6, Article 5 - § 50-f. Right of publicity}. 2020.
\newblock Civil Rights (CVR) Chapter 6, Article 5 - § 50-f. Right of publicity.
\newblock (New York State Legislature).

\bibitem[{Coffee(2023)}]{coffee2023celebrities}
Coffee, P. 2023.
\newblock Celebrities Use AI to Take Control of Their Own Images.
\newblock \emph{The Wall Street Journal}.
\newblock Using AI-generated duplicates, brands also benefit, using stars in ways they never could before.

\bibitem[{Compton and Mateas(2015)}]{compton2015casual}
Compton, K.; and Mateas, M. 2015.
\newblock Casual Creators.
\newblock In \emph{International Conference on Computational Creativity}, 228--235.

\bibitem[{{Copyright Infringement and Remedies - 17 U.S.C. § 501}(2012)}]{copyrightgov2023infringement}
{Copyright Infringement and Remedies - 17 U.S.C. § 501}. 2012.
\newblock Chapter 5: Copyright Infringement and Remedies.

\bibitem[{Coscarelli(2023)}]{coscarelli2023}
Coscarelli, J. 2023.
\newblock An A.I. Hit of Fake 'Drake' and 'The Weeknd' Rattles the Music World.
\newblock The New York Times, sec. Arts.

\bibitem[{Dayal(2022)}]{dayal2022youtube}
Dayal, M. 2022.
\newblock YouTube Increases Pressure on Instagram, TikTok to Share Ad Revenue.
\newblock 4:14 PM PDT. Photo: Chart by Shane Burke.

\bibitem[{Dhariwal et~al.(2020)Dhariwal, Jun, Payne, Kim, Radford, and Sutskever}]{dhariwal_jukebox_2020}
Dhariwal, P.; Jun, H.; Payne, C.; Kim, J.~W.; Radford, A.; and Sutskever, I. 2020.
\newblock Jukebox: {A} {Generative} {Model} for {Music}.
\newblock ArXiv:2005.00341 [cs, eess, stat].

\bibitem[{Dhariwal and Nichol(2021)}]{dhariwal2021diffusion}
Dhariwal, P.; and Nichol, A. 2021.
\newblock Diffusion Models Beat GANs on Image Synthesis.
\newblock arXiv:2105.05233.

\bibitem[{Elish(2019)}]{elish2019moral}
Elish, M.~C. 2019.
\newblock Moral crumple zones: Cautionary tales in human-robot interaction (pre-print).
\newblock \emph{Engaging Science, Technology, and Society (pre-print)}.

\bibitem[{Ellingrud et~al.(2023)Ellingrud, Sanghvi, Dandona, Madgavkar, Chui, White, and Hasebe}]{ellingrud2023generative}
Ellingrud, K.; Sanghvi, S.; Dandona, G.~S.; Madgavkar, A.; Chui, M.; White, O.; and Hasebe, P. 2023.
\newblock Generative AI and the future of work in America.
\newblock Report, McKinsey Global Institute.

\bibitem[{Epstein et~al.(2023)Epstein, Hertzmann, Herman, Mahari, Frank, Groh, Schroeder et~al.}]{epstein2023art}
Epstein, Z.; Hertzmann, A.; Herman, L.; Mahari, R.; Frank, M.~R.; Groh, M.; Schroeder, H.; et~al. 2023.
\newblock Art and the Science of Generative AI: A Deeper Dive.
\newblock arXiv:2306.04141.

\bibitem[{Epstein et~al.(2020)Epstein, Levine, Rand, and Rahwan}]{epstein2020gets}
Epstein, Z.; Levine, S.; Rand, D.~G.; and Rahwan, I. 2020.
\newblock Who gets credit for ai-generated art?
\newblock \emph{Iscience}, 23(9): 101515.

\bibitem[{Eshraghian(2020)}]{eshraghian2020human}
Eshraghian, J.~K. 2020.
\newblock Human ownership of artificial creativity.
\newblock \emph{Nature Machine Intelligence}, 2(3): 157--160.

\bibitem[{Etro and Stepanova(2018)}]{etro_power-laws_2018}
Etro, F.; and Stepanova, E. 2018.
\newblock Power-laws in art.
\newblock \emph{Physica A: Statistical Mechanics and its Applications}, 506: 217--220.

\bibitem[{Fjeld and Kortz(2017)}]{fjeld2017legal}
Fjeld, J.; and Kortz, M. 2017.
\newblock A legal anatomy of AI-generated art: Part I.
\newblock \emph{Harvard Journal of Law and Technology}.

\bibitem[{Fjeld and Kortz(2020{\natexlab{a}})}]{licencescomment2020}
Fjeld, J.; and Kortz, M. 2020{\natexlab{a}}.
\newblock {Re: USPTO Request for Comments on Intellectual Property Protection for Artificial Intelligence Innovation}.

\bibitem[{Fjeld and Kortz(2020{\natexlab{b}})}]{jointcomment2020}
Fjeld, J.; and Kortz, M. 2020{\natexlab{b}}.
\newblock {Re: WIPO Conversation on Intellectual Property (IP) and Artificial Intelligence (AI) }.

\bibitem[{Franceschelli and Musolesi(2022)}]{franceschelli2022copyright}
Franceschelli, G.; and Musolesi, M. 2022.
\newblock Copyright in generative deep learning.
\newblock \emph{Data \& Policy}, 4.

\bibitem[{Gapper(2023)}]{gapper2023}
Gapper, J. 2023.
\newblock Generative AI Should Pay Human Artists for Training.
\newblock \emph{Financial Times}.
\newblock Accessed: 2023-01-27.

\bibitem[{{Getty Images v. Stability AI}(2023)}]{gettyvstability}
{Getty Images v. Stability AI}. 2023.
\newblock In the United States District Court for the District of Delaware.
\newblock Case No. 1:23-cv-00135-UNA, Document 1, Filed 02/03/23.

\bibitem[{Gordon et~al.(2022)Gordon, Mahari, Mishra, and Epstein}]{gordon2022co}
Gordon, S.; Mahari, R.; Mishra, M.; and Epstein, Z. 2022.
\newblock Co-creation and ownership for AI radio.
\newblock In \emph{International Conference on Computational Creativity}.

\bibitem[{Gordon-Levitt(2023)}]{gordonlevitt2023ai}
Gordon-Levitt, J. 2023.
\newblock If artificial intelligence uses your work, it should pay you.
\newblock \emph{The Washington Post}.

\bibitem[{Grimes(2023)}]{grimes_twitter_2023}
Grimes. 2023.
\newblock Grimes tweet on AI 50/50 royalty sharing.

\bibitem[{Grimmelmann(2015)}]{grimmelmann2015copyright}
Grimmelmann, J. 2015.
\newblock Copyright for literate robots.
\newblock \emph{Iowa Law Review}, 101: 657.

\bibitem[{Guadamuz(2017)}]{guadamuz2017androids}
Guadamuz, A. 2017.
\newblock Do androids dream of electric copyright? Comparative analysis of originality in artificial intelligence generated works.
\newblock \emph{Intellectual Property Quarterly}.

\bibitem[{Gustar(2019)}]{gustar_fame_2019}
Gustar, A. 2019.
\newblock Fame, {Obscurity} and {Power} {Laws} in {Music} {History}.
\newblock \emph{Empirical Musicology Review}, 14(3-4): 186--215.
\newblock Number: 3-4.

\bibitem[{Gustar(2020)}]{gustar_laws_2020}
Gustar, A. 2020.
\newblock The laws of musical fame and obscurity.
\newblock \emph{Significance}, 17(5): 14--17.
\newblock \_eprint: https://onlinelibrary.wiley.com/doi/pdf/10.1111/1740-9713.01443.

\bibitem[{Heikkil\"a()}]{Greg_Rutkowski_MIT_tech_review_2022}
Heikkil\"a, M. ????
\newblock This artist is dominating {AI}-generated art. {And} he’s not happy about it.
\newblock \emph{MIT Technology Review}.

\bibitem[{Heikkilä(2022)}]{heikkiläoptout2022}
Heikkilä, M. 2022.
\newblock Artists can now opt out of the next version of Stable Diffusion.
\newblock \emph{{MIT} Technology Review}.

\bibitem[{Henderson et~al.(2023)Henderson, Li, Jurafsky, Hashimoto, Lemley, and Liang}]{henderson2023foundation}
Henderson, P.; Li, X.; Jurafsky, D.; Hashimoto, T.; Lemley, M.~A.; and Liang, P. 2023.
\newblock Foundation Models and Fair Use.
\newblock \emph{arXiv preprint arXiv:2303.15715}.

\bibitem[{Hertzmann(2018)}]{hertzmann2018can}
Hertzmann, A. 2018.
\newblock Can computers create art?
\newblock In \emph{Arts}, volume~7, 18. MDPI.

\bibitem[{Hetrick(2022)}]{dotla}
Hetrick, C. 2022.
\newblock Art Created By Artificial Intelligence Can’t Be Copyrighted, US Agency Rules.
\newblock \emph{dotLA}.

\bibitem[{Ho et~al.(2022)Ho, Chan, Saharia, Whang, Gao, Gritsenko, Kingma, Poole, Norouzi, Fleet et~al.}]{ho2022imagen}
Ho, J.; Chan, W.; Saharia, C.; Whang, J.; Gao, R.; Gritsenko, A.; Kingma, D.~P.; Poole, B.; Norouzi, M.; Fleet, D.~J.; et~al. 2022.
\newblock Imagen video: High definition video generation with diffusion models.
\newblock \emph{arXiv preprint arXiv:2210.02303}.

\bibitem[{Ho, Jain, and Abbeel(2020)}]{ho2020denoising}
Ho, J.; Jain, A.; and Abbeel, P. 2020.
\newblock Denoising Diffusion Probabilistic Models.
\newblock In Larochelle, H.; Ranzato, M.; Hadsell, R.; Balcan, M.~F.; and Lin, H., eds., \emph{Advances in Neural Information Processing Systems}, volume~33, 6840--6851. Curran Associates, Inc.

\bibitem[{Hsu and Myers(2023)}]{hsu2023can}
Hsu, T.; and Myers, S.~L. 2023.
\newblock Can We No Longer Believe Anything We See?
\newblock \emph{The New York Times}.

\bibitem[{Huang and Siddarth(2023)}]{huang2023generative}
Huang, S.; and Siddarth, D. 2023.
\newblock Generative AI and the digital commons.
\newblock \emph{arXiv preprint arXiv:2303.11074}.

\bibitem[{Karras et~al.(2018)Karras, Aila, Laine, and Lehtinen}]{karras2018progressive}
Karras, T.; Aila, T.; Laine, S.; and Lehtinen, J. 2018.
\newblock Progressive Growing of GANs for Improved Quality, Stability, and Variation.
\newblock arXiv:1710.10196.

\bibitem[{Kii(2023)}]{kii2023reflecting}
Kii, T. 2023.
\newblock Reflecting on the questions at the subcommittee of the Board of Audit and Administrative Oversight Committee.
\newblock In this article, Kii Takashi discusses his experiences and reflections on questioning during a subcommittee meeting of the Board of Audit and Administrative Oversight Committee. He focuses on issues such as the work style reform for educators and the use of generative AI in education.

\bibitem[{Kneschke(2023)}]{kneschke2023shutterstock}
Kneschke, R. 2023.
\newblock Analyse: Wie viel zahlt Shutterstock für KI-Trainingsdaten an Anbieter?
\newblock Bildagenturen, Künstliche Intelligenz, Statistik.

\bibitem[{Koh and Liang(2017)}]{koh2017understanding}
Koh, P.~W.; and Liang, P. 2017.
\newblock Understanding black-box predictions via influence functions.
\newblock In \emph{International Conference on Machine Learning}, 1885--1894. PMLR.

\bibitem[{Lanz(2023)}]{Greg_Rutkowski_Opt-out-LoRA-in_2023}
Lanz, J.~A. 2023.
\newblock Greg {Rutkowski} {Was} {Removed} {From} {Stable} {Diffusion}, {But} {AI} {Artists} {Brought} {Him} {Back}.
\newblock \emph{Decrypt}.
\newblock Section: News.

\bibitem[{Lemley and Casey(2020)}]{lemley2020fair}
Lemley, M.~A.; and Casey, B. 2020.
\newblock Fair learning.
\newblock \emph{Texas Law Review}, 99: 743.

\bibitem[{Lemley and Casey(2021)}]{lemley2021fair}
Lemley, M.~A.; and Casey, B. 2021.
\newblock Fair Learning.
\newblock \url{https://texaslawreview.org/fair-learning/}.

\bibitem[{Levendowski(2018)}]{levendowski2018copyright}
Levendowski, A. 2018.
\newblock How copyright law can fix artificial intelligence's implicit bias problem.
\newblock \emph{Washington Law Review}, 93: 579.

\bibitem[{Levin and Downes(2023)}]{levin2023regulateai}
Levin, B.; and Downes, L. 2023.
\newblock Who Is Going to Regulate AI?
\newblock \emph{Harvard Business Review}.

\bibitem[{Li and Zhang(2021)}]{li_court_2021}
Li, Q.; and Zhang, Q. 2021.
\newblock Court {Opinion} {Generation} from {Case} {Fact} {Description} with {Legal} {Basis}.
\newblock \emph{Proceedings of the AAAI Conference on Artificial Intelligence}, 35(17): 14840--14848.
\newblock Number: 17.

\bibitem[{Licensing and the Contributor~Fund(2023)}]{Shutterstock2023}
Licensing, S.~D.; and the Contributor~Fund. 2023.
\newblock Shutterstock Data Licensing and the Contributor Fund.
\newblock Accessed on: July 27, 2023.

\bibitem[{Margoni and Kretschmer(2022)}]{margoni2022deeper}
Margoni, T.; and Kretschmer, M. 2022.
\newblock A deeper look into the EU text and data mining exceptions: harmonisation, data ownership, and the future of technology.
\newblock \emph{GRUR International}, 71(8): 685--701.

\bibitem[{Metz(2022)}]{metz2022lawsuit}
Metz, C. 2022.
\newblock Lawsuit Takes Aim at the Way A.I. Is Built.

\bibitem[{Miller(2023)}]{miller2023adobe}
Miller, R. 2023.
\newblock Adobe promises artists will be compensated fairly with new generative AI product, but is fuzzy on details.
\newblock \emph{TechCrunch}.
\newblock 11:44 AM EDT.

\bibitem[{Moorhead(1995)}]{hr1295_1995}
Moorhead, C.~J. 1995.
\newblock H.R.1295 - Federal Trademark Dilution Act of 1995.
\newblock \url{https://www.congress.gov/bill/104th-congress/house-bill/1295#:~:text=Federal%20Trademark%20Dilution%20Act%20of%201995%20%2D%20Amends%20the%20Trademark%20Act,dilution%20of%20its%20distinctive%20quality.}
\newblock Sponsor: Rep. Moorhead, Carlos J. [R-CA-27] (Introduced 03/22/1995). Committees: House - Judiciary. Committee Reports: H. Rept. 104-374. Latest Action: 01/16/1996 Became Public Law No: 104-98.

\bibitem[{Murphy(2023)}]{murphy2023pay}
Murphy, B.~P. 2023.
\newblock Is There a Way to Pay Content Creators Whose Work Is Used to Train AI? Yes, but It’s Not Foolproof.
\newblock \emph{The Conversation}.

\bibitem[{Nimmer(1954)}]{nimmer1954}
Nimmer, M.~B. 1954.
\newblock The Right of Publicity.
\newblock \emph{Law \& Contemporary Problems}, 19: 203--211.

\bibitem[{O'Keefe et~al.(2020)O'Keefe, Cihon, Garfinkel, Flynn, Leung, and Dafoe}]{okeefe_windfall_2020}
O'Keefe, C.; Cihon, P.; Garfinkel, B.; Flynn, C.; Leung, J.; and Dafoe, A. 2020.
\newblock The {Windfall} {Clause}: {Distributing} the {Benefits} of {AI} for the {Common} {Good}.
\newblock In \emph{Proceedings of the {AAAI}/{ACM} {Conference} on {AI}, {Ethics}, and {Society}}, 327--331. New York NY USA: ACM.
\newblock ISBN 978-1-4503-7110-0.

\bibitem[{{Plaintiffs Richard Kadrey, Sarah Silverman, and Christopher Golden v. Meta Platforms, Inc.}(2023)}]{meta_complaint}
{Plaintiffs Richard Kadrey, Sarah Silverman, and Christopher Golden v. Meta Platforms, Inc.} 2023.
\newblock \emph{Class Action Complaint}.
\newblock United States District Court for the Northern District of California.

\bibitem[{{Plaintiffs Sarah Andersen, Kelly McKernan, and Karla Ortiz v. Stability AI Ltd., Stability AI Inc., Midjourney Inc., and DeviantArt Inc.}(2023)}]{stability_ai_complaint}
{Plaintiffs Sarah Andersen, Kelly McKernan, and Karla Ortiz v. Stability AI Ltd., Stability AI Inc., Midjourney Inc., and DeviantArt Inc.} 2023.
\newblock \emph{Class Action Complaint}.
\newblock United States District Court for the Northern District of California.

\bibitem[{Prinz(2022)}]{prinz_concentration_2022}
Prinz, A.~L. 2022.
\newblock The concentration of power in the market for contemporary art: an empirical analysis of {ArtReview}’s “{Power} 100”.
\newblock \emph{SN Business \& Economics}, 2(1): 11.

\bibitem[{Radford et~al.(2021)Radford, Kim, Hallacy, Ramesh, Goh, Agarwal, Sastry, Askell, Mishkin, Clark et~al.}]{radford2021learning}
Radford, A.; Kim, J.~W.; Hallacy, C.; Ramesh, A.; Goh, G.; Agarwal, S.; Sastry, G.; Askell, A.; Mishkin, P.; Clark, J.; et~al. 2021.
\newblock Learning transferable visual models from natural language supervision.
\newblock In \emph{International Conference on Machine Learning}, 8748--8763. PMLR.

\bibitem[{Ramalho(2017)}]{ramalho2017}
Ramalho, A. 2017.
\newblock Will Robots Rule the (Artistic) World? A Proposed Model for the Legal Status of Creations by Artificial Intelligence Systems.
\newblock \emph{SSRN Electronic Journal}.
\newblock Available at SSRN: https://ssrn.com/abstract=2987757 or http://dx.doi.org/10.2139/ssrn.2987757.

\bibitem[{Ramesh et~al.(2022)Ramesh, Dhariwal, Nichol, Chu, and Chen}]{ramesh2022hierarchical}
Ramesh, A.; Dhariwal, P.; Nichol, A.; Chu, C.; and Chen, M. 2022.
\newblock Hierarchical text-conditional image generation with clip latents.
\newblock \emph{arXiv preprint arXiv:2204.06125}.

\bibitem[{Ramirez and Hagen(2018)}]{ramirez_quantitative_2018}
Ramirez, E.~D.; and Hagen, S.~J. 2018.
\newblock The quantitative measure and statistical distribution of fame.
\newblock \emph{PLOS ONE}, 13(7): e0200196.
\newblock Publisher: Public Library of Science.

\bibitem[{Reeves and Nass(1996)}]{reeves1996media}
Reeves, B.; and Nass, C. 1996.
\newblock The media equation: How people treat computers, television, and new media like real people.
\newblock \emph{Cambridge, UK}, 10: 236605.

\bibitem[{Rombach et~al.(2022{\natexlab{a}})Rombach, Blattmann, Lorenz, Esser, and Ommer}]{Rombach_2022_CVPR}
Rombach, R.; Blattmann, A.; Lorenz, D.; Esser, P.; and Ommer, B. 2022{\natexlab{a}}.
\newblock High-Resolution Image Synthesis With Latent Diffusion Models.
\newblock In \emph{Proceedings of the IEEE/CVF Conference on Computer Vision and Pattern Recognition (CVPR)}, 10684--10695.

\bibitem[{Rombach et~al.(2022{\natexlab{b}})Rombach, Blattmann, Lorenz, Esser, and Ommer}]{rombach2022highresolution}
Rombach, R.; Blattmann, A.; Lorenz, D.; Esser, P.; and Ommer, B. 2022{\natexlab{b}}.
\newblock High-Resolution Image Synthesis with Latent Diffusion Models.
\newblock arXiv:2112.10752.

\bibitem[{Rombach, Blattmann, and Ommer(2022)}]{rombach2022textguided}
Rombach, R.; Blattmann, A.; and Ommer, B. 2022.
\newblock Text-Guided Synthesis of Artistic Images with Retrieval-Augmented Diffusion Models.
\newblock arXiv:2207.13038.

\bibitem[{Rothman(2018)}]{rothman2018right}
Rothman, J.~E. 2018.
\newblock \emph{The Right of Publicity: Privacy Reimagined for the Public World}.
\newblock Harvard University Press.
\newblock Tracing the expansion of the right of publicity beyond name and likeness, pp. 88-96.

\bibitem[{Rozansky, Antoine, and Beutler(2021)}]{rozansky2021protecting}
Rozansky, D.~A.; Antoine, H.~A.; and Beutler, J. 2021.
\newblock Protecting Image and Likeness Through Trademark Law.

\bibitem[{Saharia et~al.(2022)Saharia, Chan, Saxena, Li, Whang, Denton, Ghasemipour, Ayan, Mahdavi, Lopes et~al.}]{saharia2022photorealistic}
Saharia, C.; Chan, W.; Saxena, S.; Li, L.; Whang, J.; Denton, E.; Ghasemipour, S. K.~S.; Ayan, B.~K.; Mahdavi, S.~S.; Lopes, R.~G.; et~al. 2022.
\newblock Photorealistic Text-to-Image Diffusion Models with Deep Language Understanding.
\newblock \emph{arXiv preprint arXiv:2205.11487}.

\bibitem[{Saharia et~al.(2021)Saharia, Ho, Chan, Salimans, Fleet, and Norouzi}]{saharia2021image}
Saharia, C.; Ho, J.; Chan, W.; Salimans, T.; Fleet, D.~J.; and Norouzi, M. 2021.
\newblock Image Super-Resolution via Iterative Refinement.
\newblock arXiv:2104.07636.

\bibitem[{Samuelson(2023)}]{samuelson2023generative}
Samuelson, P. 2023.
\newblock Generative AI meets copyright.
\newblock \emph{Science}, 381: 158--161.

\bibitem[{Shan et~al.(2023)Shan, Cryan, Wenger, Zheng, Hanocka, and Zhao}]{shan2023glaze}
Shan, S.; Cryan, J.; Wenger, E.; Zheng, H.; Hanocka, R.; and Zhao, B.~Y. 2023.
\newblock Glaze: Protecting artists from style mimicry by text-to-image models.
\newblock \emph{arXiv preprint arXiv:2302.04222}.

\bibitem[{Simkin and Roychowdhury(2013)}]{simkin_mathematical_2013}
Simkin, M.~V.; and Roychowdhury, V.~P. 2013.
\newblock A {Mathematical} {Theory} of {Fame}.
\newblock \emph{Journal of Statistical Physics}, 151(1): 319--328.

\bibitem[{Singer et~al.(2022{\natexlab{a}})Singer, Polyak, Hayes, Yin, An, Zhang, Hu, Yang, Ashual, Gafni, Parikh, Gupta, and Taigman}]{singer2022makeavideo}
Singer, U.; Polyak, A.; Hayes, T.; Yin, X.; An, J.; Zhang, S.; Hu, Q.; Yang, H.; Ashual, O.; Gafni, O.; Parikh, D.; Gupta, S.; and Taigman, Y. 2022{\natexlab{a}}.
\newblock Make-A-Video: Text-to-Video Generation without Text-Video Data.
\newblock arXiv:2209.14792.

\bibitem[{Singer et~al.(2022{\natexlab{b}})Singer, Polyak, Hayes, Yin, An, Zhang, Hu, Yang, Ashual, Gafni et~al.}]{singer2022make}
Singer, U.; Polyak, A.; Hayes, T.; Yin, X.; An, J.; Zhang, S.; Hu, Q.; Yang, H.; Ashual, O.; Gafni, O.; et~al. 2022{\natexlab{b}}.
\newblock Make-a-video: Text-to-video generation without text-video data.
\newblock \emph{arXiv preprint arXiv:2209.14792}.

\bibitem[{Sobel(2017)}]{sobel2017artificial}
Sobel, B.~L. 2017.
\newblock Artificial Intelligence's Fair Use Crisis.
\newblock \emph{Columbia Journal of Law \& Arts}, 41: 45.

\bibitem[{Somepalli et~al.(2022)Somepalli, Singla, Goldblum, Geiping, and Goldstein}]{somepalli2022diffusion}
Somepalli, G.; Singla, V.; Goldblum, M.; Geiping, J.; and Goldstein, T. 2022.
\newblock Diffusion Art or Digital Forgery? Investigating Data Replication in Diffusion Models.
\newblock arXiv:2212.03860.

\bibitem[{{South Africa Companies and Intellectual Property Commission}()}]{thaler2021sa}
{South Africa Companies and Intellectual Property Commission}. ????
\newblock Acceptance of Complete Specification: Stephen Thaler's DABUS receives a patent for an AI-generated invention.

\bibitem[{{Stephen Thaler v. Comptroller General of Patents Trade Marks and Designs}(2021)}]{thaler2021}
{Stephen Thaler v. Comptroller General of Patents Trade Marks and Designs}. 2021.
\newblock Stephen Thaler v. Comptroller General of Patents Trade Marks and Designs, Case No: A3/2020/1851.
\newblock EWCA Civ 1374, Patents Court, Business and Property Courts of England and Wales.

\bibitem[{Turk and Pentland(1991)}]{turk1991eigenfaces}
Turk, M.; and Pentland, A. 1991.
\newblock Eigenfaces for Recognition.
\newblock \emph{Journal of Cognitive Neuroscience}, 3(1): 71--86.

\bibitem[{{U. S. Patent and Trademark Office}(2013)}]{uspto2013lanham}
{U. S. Patent and Trademark Office}. 2013.
\newblock The Lanham Act (Title 15 of the United States Code).
\newblock U. S. TRADEMARK LAW: FEDERAL STATUTES.
\newblock Available from U. S. Patent and Trademark Office.

\bibitem[{{Uniform Trade Secrets Act with 1985 Amendments}(1985)}]{utsa_1985}
{Uniform Trade Secrets Act with 1985 Amendments}. 1985.
\newblock \emph{Uniform Trade Secrets Act with 1985 Amendments}.

\bibitem[{{United States Patent and Trademark Office (USPTO)}(2023)}]{uspto2023trademarkinfringement}
{United States Patent and Trademark Office (USPTO)}. 2023.
\newblock About Trademark Infringement.
\newblock Sources: 15 U.S.C. §§1114, 1116-1118; Black's Law Dictionary (9th ed. 2009); McCarthy on Trademarks \& Unfair Competition (5th ed. 2017 updated quarterly).

\bibitem[{{U.S. Bureau of Labor Statistics (BLS)}(2021)}]{BLS2021}
{U.S. Bureau of Labor Statistics (BLS)}. 2021.
\newblock Projections overview and highlights, 2020–30.
\newblock \emph{Monthly Labor Review}.

\bibitem[{{U.S. Copyright Office}(2022{\natexlab{a}})}]{uscopyrightoffice_fairUse}
{U.S. Copyright Office}. 2022{\natexlab{a}}.
\newblock 17 U.S. Code § 107 - Limitations on exclusive rights: Fair use.
\newblock Circular 92.
\newblock Page.

\bibitem[{{U.S. Copyright Office}(2022{\natexlab{b}})}]{uscopyrightoffice2022}
{U.S. Copyright Office}. 2022{\natexlab{b}}.
\newblock Copyright Law of the United States and Related Laws Contained in Title 17 of the United States Code.
\newblock Circular 92.

\bibitem[{U.S. Copyright~Office(2023)}]{copyright_office_2023}
U.S. Copyright~Office, L. o.~C. 2023.
\newblock Copyright Registration Guidance: Works Containing Material Generated by Artificial Intelligence.
\newblock \url{https://www.federalregister.gov/documents/2023/03/16/2023-05321/copyright-registration-guidance-works-containing-material-generated-by-artificial-intelligence}.
\newblock Accessed: 2023-08-03.

\bibitem[{Villegas et~al.(2022)Villegas, Babaeizadeh, Kindermans, Moraldo, Zhang, Saffar, Castro, Kunze, and Erhan}]{villegas2022phenaki}
Villegas, R.; Babaeizadeh, M.; Kindermans, P.-J.; Moraldo, H.; Zhang, H.; Saffar, M.~T.; Castro, S.; Kunze, J.; and Erhan, D. 2022.
\newblock Phenaki: Variable length video generation from open domain textual description.
\newblock \emph{arXiv preprint arXiv:2210.02399}.

\bibitem[{Visser(2013)}]{visser_zipfs_2013}
Visser, M. 2013.
\newblock Zipf's law, power laws and maximum entropy.
\newblock \emph{New Journal of Physics}, 15(4): 043021.
\newblock Publisher: IOP Publishing.

\bibitem[{Wang et~al.(2023)Wang, Efros, Zhu, and Zhang}]{wang2023evaluating}
Wang, S.-Y.; Efros, A.~A.; Zhu, J.-Y.; and Zhang, R. 2023.
\newblock Evaluating Data Attribution for Text-to-Image Models.
\newblock In \emph{ICCV}.

\bibitem[{Watson(2019)}]{watson2019rhetoric}
Watson, D. 2019.
\newblock The rhetoric and reality of anthropomorphism in artificial intelligence.
\newblock \emph{Minds and Machines}, 29(3): 417--440.

\bibitem[{Williams(2023)}]{Williams2023}
Williams, K.-S. 2023.
\newblock Will.i.am wants to protect his “facial math” from AI.

\bibitem[{{World Intellectual Property Organization}(2014)}]{wipo2014monetization}
{World Intellectual Property Organization}. 2014.
\newblock \emph{Monetization of Copyright Assets by Creative Enterprises}.
\newblock Creative Industries. Geneva, Switzerland: World Intellectual Property Organization (WIPO).
\newblock ISBN 9789280524550.
\newblock License: CC BY 3.0 IGO.

\bibitem[{Yu et~al.(2022)Yu, Xu, Koh, Luong, Baid, Wang, Vasudevan, Ku, Yang, Ayan et~al.}]{yu2022scaling}
Yu, J.; Xu, Y.; Koh, J.~Y.; Luong, T.; Baid, G.; Wang, Z.; Vasudevan, V.; Ku, A.; Yang, Y.; Ayan, B.~K.; et~al. 2022.
\newblock Scaling autoregressive models for content-rich text-to-image generation.
\newblock \emph{arXiv preprint arXiv:2206.10789}.

\bibitem[{Yuan et~al.(2023)Yuan, Cun, Zhang, Li, Qi, Wang, Shan, and Zheng}]{yuan2023inserting}
Yuan, G.; Cun, X.; Zhang, Y.; Li, M.; Qi, C.; Wang, X.; Shan, Y.; and Zheng, H. 2023.
\newblock Inserting Anybody in Diffusion Models via Celeb Basis.
\newblock arXiv:2306.00926.

\end{thebibliography}
\end{document}